\documentclass[12pt,preprint]{aastex}
\usepackage{epsfig}
\usepackage{natbib}

\begin{document}
\voffset-0.5cm
\newcommand{\gsim}{\hbox{\rlap{$^>$}$_\sim$}}
\newcommand{\lsim}{\hbox{\rlap{$^<$}$_\sim$}}

\title{Origin Of The Cosmic Ray Positrons Observed Near Earth-\\
     Meson Decay Or Dark Matter Decay?}

\author{Shlomo Dado\altaffilmark{1} and Arnon Dar\altaffilmark{1}}

\altaffiltext{1}{Physics Department, Technion, Haifa 32000, Israel}

\begin{abstract} 
We show that the flux  of the high energy cosmic ray positrons 
observed near Earth is that expected from the decay of mesons produced 
by the primary cosmic rays (CRs) in the local interstellar medium.
\end{abstract}

\keywords{cosmic rays, dark matter}

\maketitle

\section{Introduction}

A wide variety of evidence points to the existence of dark matter in the 
universe which cannot be seen directly, but which can be detected by its 
gravity. Precise measurements of the angular power spectrum of the cosmic 
microwave background radiation analyzed with the standard model of 
cosmology indicate that nearly $27\%$ of the total mass-energy density of 
the universe is residing in this dark matter (Bennet et al.~2013; Ade et 
al. 2014) whose existence was first discovered in clusters of galaxies 
(Zwicky 1933,1937) and later in galaxies (Rubin 1970). The nature and 
origin of this dark matter (DM) are still unknown.

If the dark matter is made of weakly interacting massive particles (WIMPs) 
relics from the big bang, their annihilation cross section satisfies 
${\rm <\sigma_{_W}\, v>\approx 2\, \times 
10^{-26}\,cm^3/s}$ (Jungman et al.~1996), which is 
typical of weakly interacting $\sim$TeV-mass particles. Annihilation/decay 
of such particles can produce stable particles such as electrons, 
positrons, protons, antiprotons, neutrinos and gamma rays, whose energy 
cannot exceed the mass of the DM particles. Hence, in recent years, bumps 
with a sharp cutoff due to the decay and/or annihilation of WIMPs
were looked for in high precision measurements of the energy
spectrum of such cosmic ray (CR) particles.
A bump around E=620 GeV with a sharp cutoff around E=800 GeV in the 
combined flux ${\rm \Phi_{e^\pm}}$  of ${\rm e^\pm}$ cosmic rays (CRs) was 
reported by the Advanced Thin Ionization Calorimeter (ATIC) balloon 
experiment (Chang et al.~ 2008). It was 
interpreted as a possible dark matter signal 
(e.g., Bergstrom et al.~2008; Cholis et al.~2009; Cirelli et 
al.~2009; Arkani-Hamed et al. 2009; Pohl 2009). However, such a 
peak was not confirmed by measurements with the ground based Cherenkov 
Telescopes of High Energy 
Stereoscopic System (H.E.S.S.) (Aharonian et al.~2009), and later 
with more precise measurements with the Large Area Telescope (LAT) aboard 
the Fermi satellite (Ackermann et al.~2010) and with 
the Alpha Magnetic Spectrometer (AMS) on
the International Space Station (Aguilar et al.~2014).
But, at the same time, the energy spectrum of ${\rm e^\pm}$ CRs 
measured with H.E.S.S. (Aharonian et al.~2008) showed a sharp break
near 1 TeV whose origin is still not clear.  

Shortly after the ATIC report (Chang et al.~2008), an unexpected increase 
of the CR positron fraction ${\rm \Phi(e^+)/( \Phi(e^+)+ \Phi(e^-))}$ as 
a function of energy between $\sim$10 GeV and $\sim$100 GeV, observed with 
the Payload for Antimatter Matter Exploration and Light-nuclei 
Astrophysics (PAMELA) satellite was reported (Adriani et al. 
2009,2010,2013). Unlike the ATIC bump, the rise with energy of the 
positron fraction has been confirmed by measurements of the separate 
CR ray fluxes of ${\rm e^{\pm}}$ with Fermi-LAT (Ackermann et al. 2012), 
and with AMS (Aguilar et al.~2013,2014; Accardo et al.~2014) with much 
larger statistics and smaller systematic errors. In the high precision AMS 
data, this rise continues beyond 100 GeV but  with a decreasing rate that 
appears to level off around $275\pm 32$ GeV (Accardo et al.~2014, Aguilar 
et al. 2014, Pohl 2014).

The measured flux of CR positrons and the increasing positron fraction 
were in stark contrast with those predicted for secondary production of 
positrons in the interstellar medium (ISM) by primary CRs from detailed 
calculations with the elaborate GALPROP code (e.g., Moskalenko and Strong 
1998a,b). This sparked the publication of several alternative 
explanations to the origin of the excess of high energy CR ${\rm e^+}$
flux, such as dark matter (e.g., Bergstrom et al. 2008,2009,2013; 
Bergstrom 2009; for a recent review see, e.g., 
Ibarra et al. 2013 and references therein), positron emission from a few 
nearby pulsars and/or supernova remnants (e.g., Hooper et al. 2009; Shaviv 
et al. 2009), secondary production of positrons in collisions of primary 
CRs in/near their sources (Dado \& Dar~2010) such as the highly 
relativistic jets launched mainly by supernovae Type Ic and acreting 
stellar mass and massive black holes (e.g., Dar and De R\'ujula 2004, 2008 
and references therein) and/or the spherical ejecta in supernova remnants 
(SNRs) (Blasi 2009; Stawarz et al. 2010). Other authors (e.g., Blum et 
al.~2013; Cowsik et al.~ 2014) maintained that secondary production of 
positrons in the ISM by the primary CRs can explain the observed flux of 
CR positrons although their calculations involved incorrect assumptions 
concerning the energy-loss of positrons in the ISM.

The reported isotropy of the flux of high energy CR positrons observed 
with AMS (Accardo et al.~2014), however, is in tension with the assumption 
that the observed flux of ${\rm e^+}$ CRs was produced by a few relatively 
nearby pulsars (e.g., Linden and Profumo 2013) or supernova remnants. At 
the same time, the high precision data of AMS02 (Aguilar et al. 2014) on 
the combined ${\rm e^\pm}$ flux up to 1 TeV and the separate ${\rm e^-}$ 
and ${\rm e^+}$ fluxes, and the positron fraction (Accardo et al.~2014, 
Aguilar et al.~2014) that appear to flatten above 200 GeV, do not show yet 
any convincing evidence for a cutoff/sharp decline, which begins below 
$\sim 700$ GeV that could be associated with a decay/annihilation of dark 
matter, particles.

In this paper, we re-examine the possibility that the main origin of the 
high energy CR positrons observed near Earth is secondary production in 
hadronic collisions of the primary cosmic ray protons and nuclei in the 
ISM (Blum et al.~2013; Cowsik et al.~2014). We focus on the high energy 
(E$>$ 10 GeV) behavior of the flux of CR ${\rm e^+}$'s where it may be 
dominated by the decay/annihilation of dark matter particles. We consider 
in detail, the energy-loss of ${\rm e^+}$'s above 10 GeV where it is 
dominated by synchrotron radiation, inverse Compton scattering, triple 
pair production, and escape from the Galaxy by diffusion in its turbulent 
magnetic fields. (Other energy loss mechanisms that are important only at 
energy well below 10 GeV, such as Coulomb scattering, ionization and 
bremsstrahlung, as well as threshold effects, geomagnetic shielding and 
solar modulation, are not considered in detail, but are included for 
completeness through a best fit phenomenological depletion factor ${\rm 
D_{e^\pm}(E)=1-exp(-(E/V)^\alpha)}$, which depends on the time of the 
measurements and on the elecctric  charge ($\pm$) 
but does not affect the behavior at E$>$10 GeV.)
Using only standard particle physics, the steady state 
approximation for CR propagation, and the measured properties of the 
primary CRs and the local Galactic environment, 
we show that the spectrum 
of high energy CR ${\rm e^+}$'s measured by AMS (Aguilar et al.~2014) 
with high statistics and small systematic errors is consistent with that 
expected from secondary production of positrons in hadronic interactions 
of the primary CRs in the local ISM. In particular, the hardening of their 
spectrum with increasing energy between $\sim$ 30 and 150 GeV and its 
softening beyond 250 GeV, as observed with AMS02 (Aguilar et al.~2014), 
are expected from the transition of their inverse Compton scattering of  
diffuse Galactic light from the Thomson regime to the Klein-Nishina 
regime (Schlickeiser \& Ruppel~2010).

Moreover, the break near 1 TeV in the combined ${\rm e^\pm}$ flux, 
suggested by the data from H.E.S.S. (Aharonian et al.~2008) may be the 
maximum re-acceleration energy of electrons in the ISM due to their 
radiative energy-loss and escape from the Galaxy by diffusion. Indeed a 
break was reproduced in numerical calculations (e.g., Stawarz et 
al.~2010), which have used the elaborate GALPROP numerical code 
(Moskalenko \& Strong 1998a,b). Other standard astroparticle physics
interpretations 
rather than a dark matter annihilation/decay signal are shortly discussed.

\section{CR Production of positrons in the local ISM:}
The flux of primary Cosmic ray nucleons (free protons and nucleons bound 
in atomic nuclei) in the energy range between several GeV and PeV 
per nucleon observed near Earth is well 
described by (e.g., Olive et al.~2014), 
\begin{equation}
{\rm \Phi_p(E)\approx  1.8\,(E/GeV)^{-\beta}\,\, fu },
\label{Eq.1}
\end{equation}
where ${\rm \beta =2.7}$ and ${\rm fu=1/(GeV\, cm^2\, s\,\, sr)}$ 
is the flux unit. The contribution of nucleons 
bound in atomic nuclei (A,Z) to the cosmic ray flux at the same energy per 
nucleon is less than $5\%$ and comes mainly from light nuclei whose 
inelastic cross section per nucleon is ${\rm \sim\sigma_{pA}/A\approx 
\sigma_{pp}}$, i.e., roughly that of free protons, while their 
effective rigidity (R=AE/Z) is twice than that of free protons. 
Hence, due to Feynman scaling (1969), secondary 
production of high energy positrons through the decay of ${\rm\pi}$ and 
${\rm K}$ mesons, which are produced in hadronic collisions of cosmic ray 
nucleons with the baryons in the local ISM  
where the mean baryon density is ${\rm n_{ism}\approx 0.9\, cm^{-3}}$
(Kalberla \& Dedes, 2008) inject into 
the CR halo a flux of ${\rm e^+}$'s per unit volume roughly  at a rate
\begin{equation}
{\rm J_{e^+}\approx K_{e^+}\, \sigma_{in}(pp)\, c\,n_{ism}\, \Phi_p(E)}\,. 
\label{Eq.2}
\end{equation}  
where ${\rm K_{e^+}}$ is a constant 
that depends only on the power-law index of the product 
${\rm \sigma_{in}(pp)\, \Phi_p(E)}$ but not on E itself. 
In a steady state, the flux ${\rm \Phi_{e^+}(E)}$  of secondary 
${\rm e^+}$'s produced in the ISM  satisfies 
\begin{equation}
{\rm {d\over dE}[b(E)\,\Phi_{e^+}(E)]=J_{e^+}(E)}
\label{Eq.3}
\end{equation} 
where ${\rm b(E)=dE/dt}$ is the loss rate of ${\rm e^+}$ energy
by radiation (rad) and by escape (esc) from the Galaxy by diffusion 
through its turbulent magnetic field.  
The  solution of Eq.~(3) for ${\rm \Phi_p\propto E^{-\beta}}$ is
\begin{equation}
{\rm \Phi_{e^+}(E)\approx  K_{e^+}\,\sigma_{in}(pp)\,n_{ism}\,c\, 
\tau_e\,\Phi_p/(\beta_j-1)}\,.
\label{Eq.4}
\end{equation}
where ${\rm \sigma_{in}(pp)\approx 30\, (E/GeV)^{0.06}}$  mb, 
${\rm K_{e^+}\approx 7\times 10^{-3}}$ for 
${\beta_j\approx\beta-0.06=2.64}$, 
and ${\rm \tau_e=E/(dE/dt)}$ is the mean life-time
of positrons in the ISM due to their escape (esc) from the Galaxy by 
diffusion and radiative (rad) energy losses
(inverse Compton scattering of background photons and synchrotron radiation). 
Hence, the expected flux 
of high energy CR ${\rm e^+}$'s near Earth is simply
\begin{equation}
{\rm \Phi_{e^+}(E)\approx  6.22\times 10^{-18}\,E^{-2.64}\, \tau_e(E)}\, 
fu.
\label{Eq.5}
\end{equation} 
where the combined radiative losses and escape by diffusion 
yields 
\begin{equation}
{\rm 1/\tau_e= 1/\tau_{esc}+1/\tau_{rad}}\,.
\label{Eq.6}
\end{equation}
At high energy, the rigidities  of ${\rm e^+}$'s and protons become 
practically equal. Consequently,  their escape  
times by diffusion become practically equal. 
For random Galactic magnetic fields
with  a Kolmogorov spectrum 
\begin{equation}
{\rm \tau_{esc}\approx (7.5\pm 1.5) \times 10^{14}\, (E/GeV)^{-1/3}\, s}
\label{Eq.7}
\end{equation}
where the normalization has been adjusted to the value obtained 
by Lipari (2014) from a leaky box model analysis of
the flux ratio  ${\rm ^{10}Be/^9Be}$ measured
with  the Cosmic Ray Isotope Spectrometer (CRIS)  in 
the energy range  70-145 MeV/nucleon (Yanasak et al.~2001). 
At the CR ankle near ${\rm E\approx 4\times 10^9\, GeV}$,
this normalization yields
${\rm \tau_{esc}=(15\pm 3)}$ ky, consistent with 
an expected free escape time H/c=$5\pm 1$ kpc/c=$(16.3\pm 3.2)$ ky,  
where H is the typical scale height of the Galactic cosmic
ray halo perpendicular to the Galactic disk 
(e.g., Trotta et al.~2011).

Synchrotron emission and inverse Compton scattering of background
photons in the Thomson regime yield 
\begin{equation}
{\rm \tau_{rad}\approx  {3\,(m_e\,c^2)^2\over
4\,\sigma_{_T}\, c\, U\, E} \approx 0.67\times 10^{16}\, (E/GeV)^{-1}\, s}
\label{Eq.8}
\end{equation}
where ${\rm \sigma_{_T}\approx}$0.66 barn is the Thomson cross section 
and ${\rm U\sim 1.47\, eV/cm^3}$  is the energy density  
of the background radiations  plus the magnetic field
in the local ISM. Roughly, the magnetic field ${\rm B\sim 4\, \mu\, G}$
in the local ISM contributes  ${\rm B^2/8\,\pi\sim 0.40\, eV/cm^3}$ to U,
the diffuse Galactic light (DGL) contributes ${\rm \sim 0.39\,eV/cm^3}$,
the  far infrared (FIR) light contributes ${\rm \sim 0.40\, eV/cm^3}$,
and the cosmic microwave background (CMB) 
contributes  ${\rm 0.26\, eV/cm^3}$ (e.g., Porter et al.~2006;
Schlickeiser \& Ruppel 2010).

In the Klein-Nishina regime,
(${\rm w=4\, E_e\, \epsilon_\gamma/ (m_e\,c^2)^2 \gsim  1}$),
the energy loss-rate by ICS 
is given by (e.g., Blumenthal \& Gould~1970)
\begin{equation} 
{\rm {dE\over dt}\approx -\,{3\, \sigma_{_T}\, c\, U_\gamma\,
(m_e\,c^2)^2\over 8\, \epsilon_\gamma^2}[ln w-11/6]\,.}
\label{Eq.9}
\end{equation}
Consequently, as can be seen from Eq.~(3), in a steady state,
an injected power-law spectrum of electrons 
becomes softer in the Thomson regime,
${\rm E^{-\beta_j}\rightarrow E^{-\beta_j-1}}$, whereas in the 
Klein-Nishina
regime it becomes harder,  ${\rm E^{-\beta_j}\rightarrow E^{-\beta_j+1}}$.

In the case of a gray body radiation of temperature T,
the transition from the Thomson regime to the Klein-Nishina regime
can be interpolated through replacing U(E) by 
\begin{equation}
{\rm U_{KN}(E)\approx U(E)\,{E_{KN}^2/(E^2+E_{KN}^2})}
\label{Eq.10}
\end{equation}
where (Schlickeiser \& Ruppel 2010)
\begin{equation}
{\rm E_{KN}\approx {0.27\, (m_e\,c^2)^2\over k\, T}}\,,
\label{Eq.11}
\end{equation}
with k being  the Boltzman constant. For the local DGL, T$\approx 
5700^o$K and ${\rm E_{KN}\approx 140\, GeV}$. For  the Galactic FIR 
radiation and the CMB, ${\rm E_{KN}}$ is  well above 10 TeV.

Since the estimated values of the local energy densities of the DGL and 
FIR, as well as the local B suffer from uncertainties, we have best fitted 
their values within the  errors of their estimated values. At 100 
GeV, they yield for instance, ${\rm \tau_{esc}=1.62\times 10^{14}\,s}$, 
${\, \tau_{rad}\approx 7.2\times 10^{13}\, s}$, and consequently 
${\rm \tau_e\approx 0.50\times 10^{14}\, s\approx 1.6\, MY}$.

The flux of high energy ${\rm e^+}$ CRs measured with 
AMS02 (Aguilar et al.~2014) and the flux expected from CR production of 
${\rm e^+}$'s in the ISM as given by Eq.~(5) multiplied by the low-energy 
best fit phenomenological depletion factor 
${\rm D_{e^+}(E)=1-exp(-(E/V)^\alpha)}$ with V=3.90 GeV and 
${\rm \alpha=1.78}$ are compared in Fig.~1.

\section{Origin of the CR ${\rm e^-}$ flux}
Most of the  Galactic accelerators of high energy cosmic rays 
(supernova remnants, pulsars, gamma ray bursts, and the massive black hole 
near the Galactic  center) are located in the Galactic disk
within $\sim~4$ kpc from its center (e.g., Case \&  Bhattacharya~1998;
Yusifov, \& I. Kucuk~2004; Lorimer et al.~2006).
Presumably, the Fermi acceleration of high energy 
electrons, protons and nuclei is simultaneous and imparts 
to them  the same Lorentz factor distribution (e.g. Dar \& De 
R\'ujula~2008), i.e., a source  distribution  ${\rm J_{e^-}\propto J_p}$.
These primary CRs can reach Earth from their common sources by diffusion 
(dif) through the turbulent Galactic magnetic fields. 
High energy CR protons and electrons have practically 
the same rigidity (except for the sign). Consequently,    
without energy losses,  they would  have reached 
Earth from their common sources after the same mean diffusion time,
and their fluxes in the local ISM then
would have satisfied ${\rm \phi_{e^-}(E)\propto \Phi_p(E)}$.
This relation can hold as long as their energy
loss by radiation is quite small during their diffusion time 
from source to Earth.

The radiative energy loss of ${\rm e^-}$ CRs in the ISM 
satisfies ${\rm dE/dt=-b\,E^2}$ whose solution 
is ${E=E_0\,[1-b\,E\,t_{dif}]}$ where E is the observed energy of the 
${\rm e^-}$ CR near Earth and ${\rm E_0}$ is its injection energy. It 
implies that most of the sources that are located beyond $\sim 4$ kpc 
cannot contribute much to the observed local flux of CR ${\rm e^-}$', 
unless their CRs are re-accelerated in the ISM or transported 
ballistically 
over long Galactic distances by highly relativistic jets, such as those 
emitted in supernova explosions of Type Ic, most of which do not point in 
our direction (e.g., Dar \& Plaga~1999, Dar \& De R\'ujula~2008).

If the escape time of CR protons and electrons 
from their sources into the ISM 
is much shorter than their energy loss rate there, 
then the Fermi accelerated
high energy electrons are injected  into the ISM at a rate
${\rm J_e\propto \Phi_p/\tau_{esc}(E)}$.
Hence, their steady state flux near Earth satisfies
\begin{equation}
{\rm \Phi_{e^-}= D_{e^-}(E)\, A_{e^-}\,E^{-\beta}\,\tau_e/\tau_{esc}
+0.74\, \Phi_{e^+}}\,, 
\label{Eq.12}
\end{equation}
The first term on the right hand side (RHS) of Eq.~(12) is due to primary 
CR ${\rm e^-}$'s, presumably re-accelerated in the ISM 
or transported ballistically  by highly relativistic jets from all distant 
sources together with the CR protons (${\rm A_{e^-}}$ is an unknown 
flux normalization constant, which is  treated as a free parameter.)
The second term on the  RHS  is the contribution 
from hadronic production of high energy ${\rm e^-}$'s in pp collisions,
in the local ISM assuming 
an inclusive  production ratio identical to that of muons,   
${\rm e^-/e^+\approx \mu^-/\mu^+=0.74}$, which was measured for 
atmospheric 
muons and in accelerator experiments
(Haino et al. 2004; Archard et al. 2004; Adamson et al. 2007;
Khachatryan et al. 2010; Agafonova et al. 2010). 
Eq.~(12) can also be written as 
\begin{equation}
{\rm E^3 \Phi_{e^-}= {D_{e^-}(E)\, A_{e^-}\,E^{3-\beta}\over
(1+ \tau_{esc}/\tau_{rad})}+0.74\,E^3\, \Phi_{e^+}}\,,
\label{Eq.13}
\end{equation}
Assuming that the mean radiation field in the CR halo can be represented 
by that in the solar neighborhood, the predicted flux of CR ${\rm e^-}$'s
as given by Eq.~(13) for the best fit values 
${\rm A_{e^-}=0.27\, GeV^3\, fu}$, V=2.25 GeV and ${\alpha=1.46}$ 
is compared in Fig.~2 to
the CR ${\rm e^-}$ flux measured with AMS02 (Aguilar et al.~2014).
The agreement is quite satisfactory (${\rm \chi^2/df=0.85}$).

\section{The ${\rm e^+}$ fraction and the combined ${\rm e^\pm}$ flux} 
The positron fraction 
${\rm \Phi_{e^+}/\Phi_e}$ as a function of energy obtained
from our predicted ${\rm e^\pm}$ fluxes, which are 
plotted in Figs.~1 and 2, and the positron fraction
measured with AMS02 (Accardo et al.~2014) near Earth are compared in 
Fig.~3 (${\rm \chi^2/df=0.35}$). The combined ${\rm e^\pm}$ flux 
${\rm \Phi_e= \Phi_{e^+}+\Phi_{e^-}}$
measured near Earth with AMS02 (Aguilar et al. 2014) and 
the expected flux obtained from standard astroparticle physics
are compared in Fig.~4 (${\rm \chi^2/df=1.27}$). As can be seen from
Figs. 1-4, the agreement between the near Earth ${\rm e^\pm}$ CR fluxes
measured with AMS02 and those expected from
standard astroparticle physics is  quite good. 
 
\section{Dark matter annihilation signal in AMS02 ?}
The AMS collaboration presented  an impressive  'minimal model' fit
(${\rm \chi^2/df=0.63}$)
to their updated data (Accardo et al.~2014) on the CR positron fraction,
which is shown in Fig.~5. In this model the ${\rm e^\pm}$ CR 
fluxes were parametrized as the sum of a
power law spectrum and a common source
term with an exponential cutoff, which may represent a dark 
matter decay/annihilation contribution,
\begin{equation}
{\rm \Phi_{e^+}=C_{e^+}\,E^{-\gamma_{e^+}}+C_s\, e^{-E/E_s}\,,}
\label{Eq.14}
\end{equation}
\begin{equation}
{\rm \Phi_{e^-}=C_{e^-}\,E^{-\gamma_{e^-}}+C_s\, e^{-E/E_s}\,,}
\label{Eq.15}
\end{equation}
with E in GeV.
However, the  positron fraction ${\rm \Phi_{e^+}/(\Phi_{e^-}+\Phi_{e^+})}$
is invariant under division of both fluxes   
by an arbitrary E-dependent function. In particular,
by dividing the minimal model  fluxes by ${\rm C_{e^+}\,E^{-\gamma_{e^+}}}$,
a very good  best fit ($\chi^2/df=36.4/58$)
to the positron fraction in the energy range  1 to
500 GeV yielded (Accardo et al.~2014) ${\rm C_{e^+}/C_{e^-}=0.091\pm 0.001}$,
${\rm \gamma_{e^+}-\gamma_{e^-}=0.56\pm 0.03}$,
${\rm C_s/C_{e^-}=0.0061\pm 0.0009}$,
${\rm \gamma_s-\gamma_{e^-}=-0.72 \pm 0.04}$ and
a cutoff parameter ${\rm E_s=0.54\pm 0.17}$ TeV.
These best fit parameters  specify the ${\rm e^\pm}$ fluxes
of the minimal model only up to an unknown common power-law factor
${\rm C_{e^+}E^{-\gamma_{e^+}}}$.

A major feature of the minimal model is the sharp decline of the positron 
fraction above $\sim$500 GeV. 
In order to test the AMS minimal model, we have best fitted ${\rm 
\Phi_{e^+}}$ and ${\rm \Phi_{e^-}}$ as given by Eqs. (13) and (14), 
respectively, to the separate ${\rm e^+}$ and ${\rm e^-}$ fluxes measured 
with AMS02 (Aguilar et al. 2014). The  best fit parameters were 
constrained to reproduce the above relations between the 
parameters of the  minimal model. Unlike the 
minimal model impressive fit to the positron fraction 
published by the AMS collaboration, 
(Accardo et al.~2014) as shown in Fig.~5, the best fits to the separate 
${\rm e^+}$ and ${\rm e^-}$ fluxes, and to their sum, are rather poor, 
as shown in Figs.~6-8. That is true in particular for E$<10$ GeV 
and E$>300$ GeV.

One may argue, however, that the minimal model represents well the source 
spectra of $e^{\pm}$ CRs, which are later modified by propagation effects 
in/near source, in the ISM and in the heliosphere. But, because solar 
modulation, escape by diffusion in the turbulent Galactic magnetic fields, 
and energy-loss through ionization, bremsstrhalung, synchrotron radiation, 
inverse Compton scattering and pair production, which strongly affect the 
observed spectra of $e^\pm$ CRs near Earth, are independent of the sign of 
charge, they modify the injection spectra of $e^\pm$ by the same 
energy-loss factor ${\rm E/(dE/dt)=b(E)}$. However, the spatial 
distribution of Galactic dark matter is quite different from that of the 
more conventional sources of primary ${\rm e^{-}}$ and secondary ${\rm 
e^\pm}$ CRs. This makes rather unlikely the possibility that the strong 
modifications of the source spectra of both high energy ${\rm 
e^+}$ and ${\rm e^-}$ CRs observed near Earth are nearly identical.

\section{CR ${\rm e^\pm}$ knee near TeV?} 

The energy spectrum of high energy ${\rm e^\pm}$ CRs measured with 
H.E.S.S. (Aharonian et al.~2008,2009) suggests a sharp break in the 
combined ${\rm e^\pm}$ spectrum near E$\sim$TeV. This is shown in Fig.~10 
where we plotted a smooth cutoff power-law fit to the combined ${\rm 
e^\pm}$ flux measured with AMS02 (Aguilar et al.~2014) and with H.E.S.S 
(Aharonian et al 2008, 2009) after normalizing the H.E.S.S data within 
their reported systematic errors to match the more precise AMS02 sub-TeV 
data. This fit yield a break energy around 1 TeV.

Fermi acceleration of electrons is cut-off by synchrotron emission 
when the energy loss rate by synchrotron radiation exceeds the  
energy gain by magnetic deflections. This synchrotron cutoff energy 
is given roughly by ${\rm E\approx \sqrt{6\, 
e/\sigma_{_{T}}\, B}\, m_e\, c^2}$. However, the typical magnetic 
field B in supernova remnants (SNRs), which are widely accepted as the 
main source of high energy Galactic CR electrons, is well below 100 $\mu$G.
The corresponding synchrotron cutoff in the energy spectrum of CR 
${\rm e}$'s accelerated in SNRs is expected at $\gsim$ 1 PeV, well above 
that suggested by the H.E.S.S. data. 

The observed flux of  primary ${\rm e^-}$ CRs is also cut-off  
when their  radiative life becomes shorter than their travel time
by diffusion from source to Earth as discussed in Section 3.  However,
for the main Galactic sources, this cut-off is much below the H.E.S.S. 
break/cut-off.

The H.E.S.S. cutoff could be a dark matter signal. However, the data of 
Fermi-LAT (Ackermann et al.~2010) and AMS02 (Aguilar et al.~2014) 
published so far, do not show any hint for such a break below TeV.  
Moreover, such a break, if real, still could be explained by standard 
astroparticle physics rather than as a dark matter signal. The origin of a 
break could be, e.g.,

\noindent 
(A) A re-acceleration cutoff when the re-acceleration time in the 
Galactic ISM exceeds the electrons' life time due to 
their radiative energy losses and  escape from the Galaxy by diffusion.    

\noindent
(B) CR ${\rm e^-}$ "knee" at ${\rm E_{knee}(e^-)\!=
\!(m_e/m_p)E_{knee}(p)\!\approx\! 1\, TeV}$ in the spectrum
of ${\rm e^-}$ CRs, which are Fermi accelerated together 
with protons and nuclei by the highly 
relativistic jets launched in  SNe Ic 
to the same Lorentz factor distribution 
(e.g., Dar and De R\'ujula~2008). (${\rm E_{knee}(p)\approx 2}$ PeV 
is the CR "knee" in the flux {\it per nucleon} of CR nuclei).

(C)  A cutoff when the energy deposition rate of high energy electrons in 
the ISM by highly relativistic jets cannot compensate anymore their energy 
loss by radiation and escape from the Galaxy by diffusion.

For other possible standard astroparticle physics origins of the H.E.S.S 
break see, e.g., Stawarz et al. (2010) and Cowsik et al. (2014).

\section{Conclusions}
Figs. 1-4 demonstrate that standard astroparticle physics, such as
positron production in the ISM in hadronic collisions of the primary 
CRs with gas in  the local ISM and electron acceleration 
in a variety of Galactic sources, can explain the observed fluxes 
of CR electrons and positrons near Earth, which were measured 
with  AMS02 (Accardo et al. 2014; Aguilar et al. 2014) without 
invoking a contribution from annihilation/decay of dark matter particles. 
In particular, the hardening of the positron flux between 20 and 200 GeV
can be due to the Klein-Nishina supression of the energy loss of 
positrons by inverse Compton scattering from the diffuse Galactic light
in the local ISM.  

Neither the current published data on the spectra of the separate and 
combined ${\rm e^\pm}$ CRs near Earth, nor the minimal model fit by the 
AMS collaboration to their measured positron fraction (Accardo et 
al.~2014) provide  compelling  evidence or  hints for a contribution from 
decay or annihilation of dark  matter particles. 

If hadronic production 
of mesons in the local ISM
at energies well above TeV becomes the main source of 
high energy ${\rm e^\pm}$ CRs near Earth, then the positron fraction 
should reach there the  asymptotic value $\approx 0.57$ (Dado \& Dar 
2010; Cowsik et al.~2010) expected from the observed charge ratio
${\rm \mu^+/\mu^-\approx 1.35}$  of high energy atmospheric  
and accelerator muons.

The existence of a break near TeV in the combined ${\rm e^\pm}$ flux 
measured with H.E.S.S., still needs confirmation from independent and more 
precise measurements with instruments such as Fermi-LAT and the AMS. If 
verified, standard astroparticle physics may still explain it
as shortly discussed in Section 6. Moreover, 
any alternative interpretation of its origin, including dark matter, must
provide falsifiable predictions to establish its validity.

\begin{figure}[]
\centering
\epsfig{file=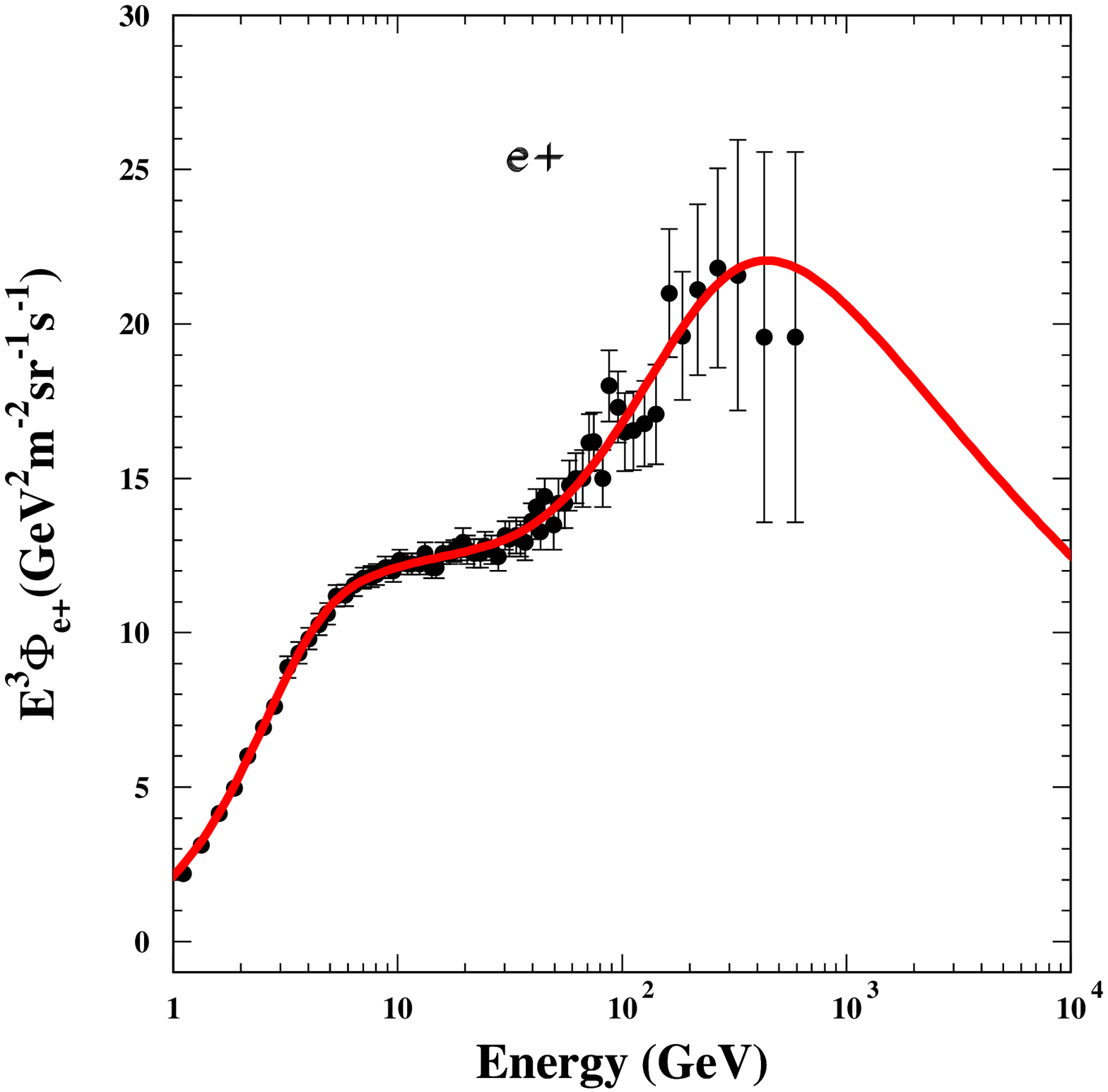,width=16.cm,height=16.cm}
\caption{Comparison between the high energy ${\rm e^+}$ CR flux 
measured with AMS02 (Aguilar et al.~2014) and the secondary ${\rm e^+}$ 
flux  expected from hadronic interactions of the primary CR nucleons 
in the local ISM.}
\label{Fig1}
\end{figure}

\begin{figure}[]
\centering
\epsfig{file=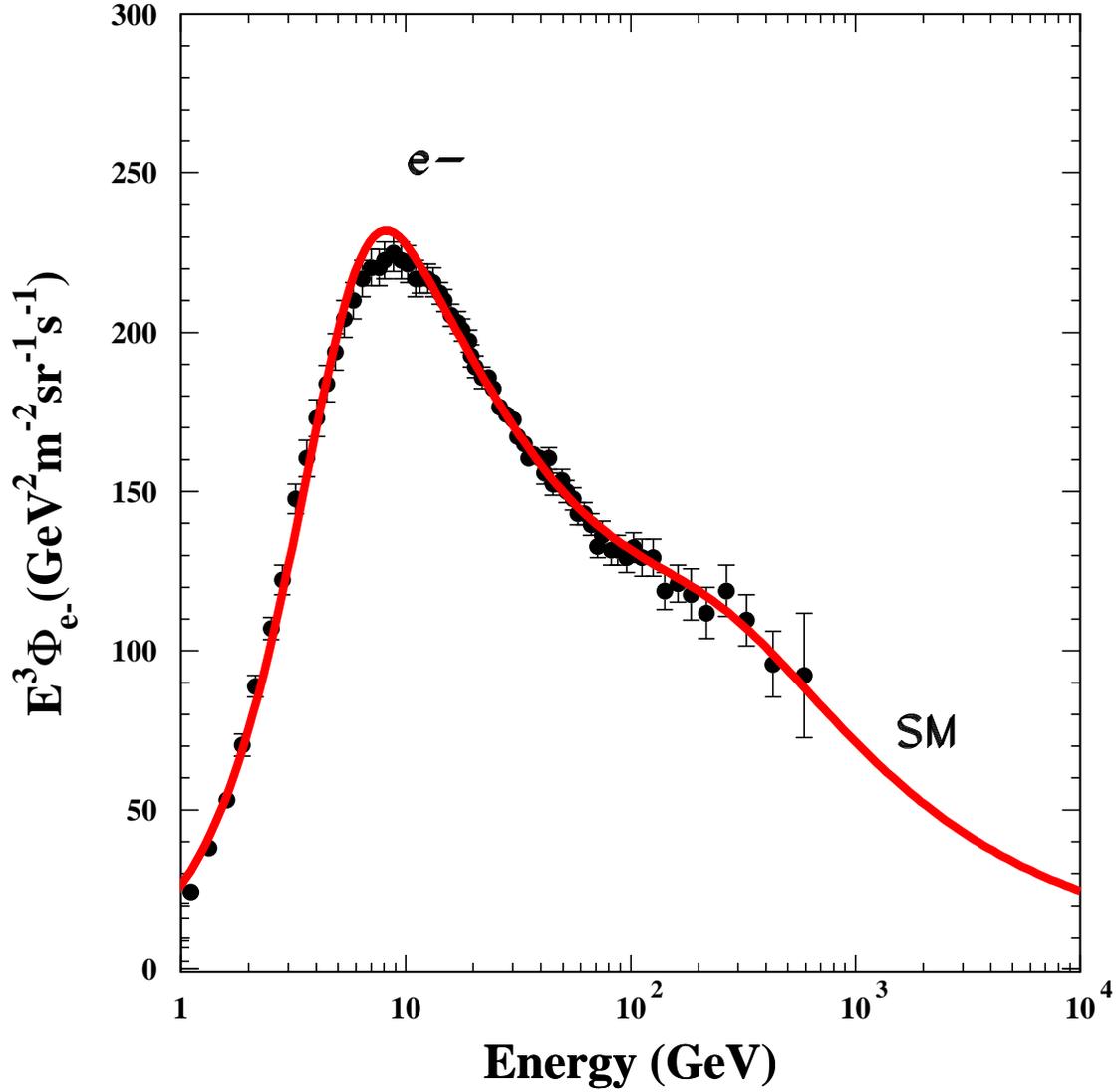,width=16.cm,height=16.cm}
\caption{Comparison between the flux of high energy ${\rm e^-}$ CRs
measured with AMS02 (Aguilar et al.~2014) and the flux 
expected from  Fermi acceleration in source of   ${\rm e^-}$'s plus  
secondary production  in/near source and  in the ISM.}
\label{Fig2}
\end{figure}

\begin{figure}[] 
\centering 
\epsfig{file=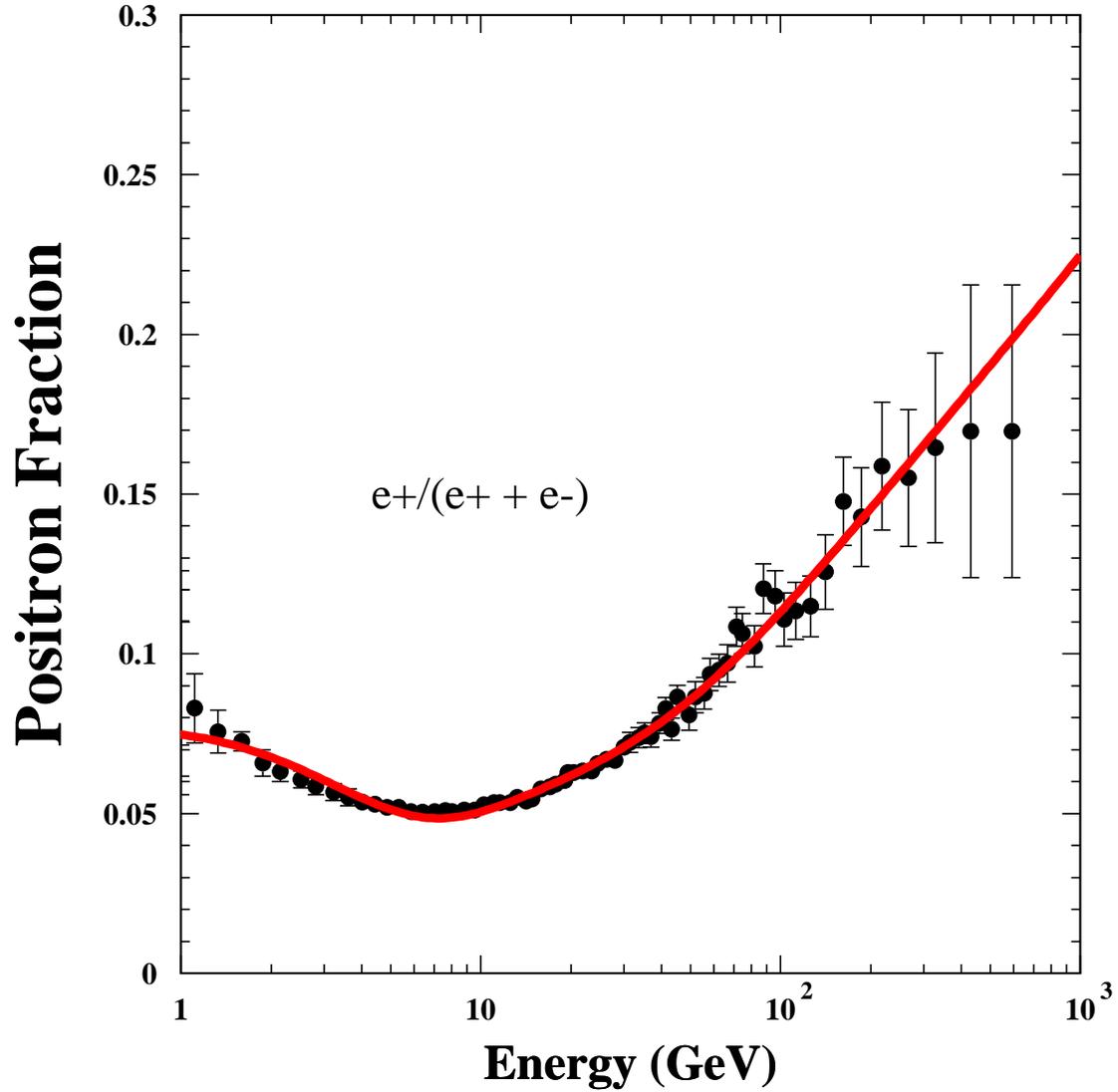,width=16.cm,height=16.cm} 
\caption{Comparison between the CR positron fraction near Earth as   
measured with AMS02 (Accardo et al.~2014) and the positron fraction 
obtained from the predicted ${\rm e^\pm}$ CR fluxes shown in Figs.~1 and 2.} 
\label{Fig3}
\end{figure}

\begin{figure}[]
\centering
\epsfig{file=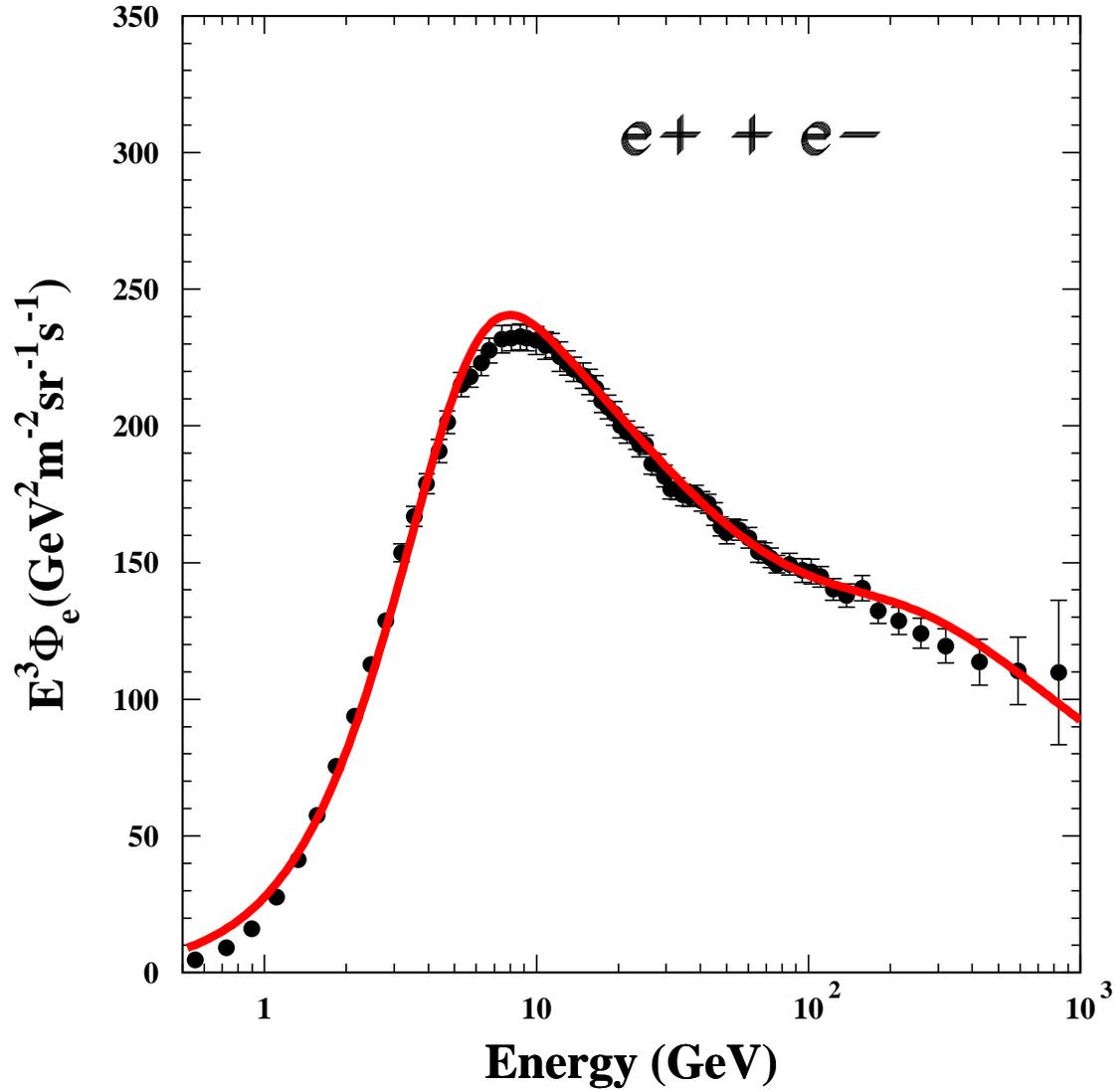,width=16.cm,height=16.cm} 
\caption{Comparison between the combined
${\rm e^\pm}$ CR flux measured with AMS02  
(Aguilar  al.~2014) and the flux obtained 
from our standard particle astrophysics model.} 
\label{Fig4}
\end{figure}

\begin{figure}[]
\centering
\epsfig{file=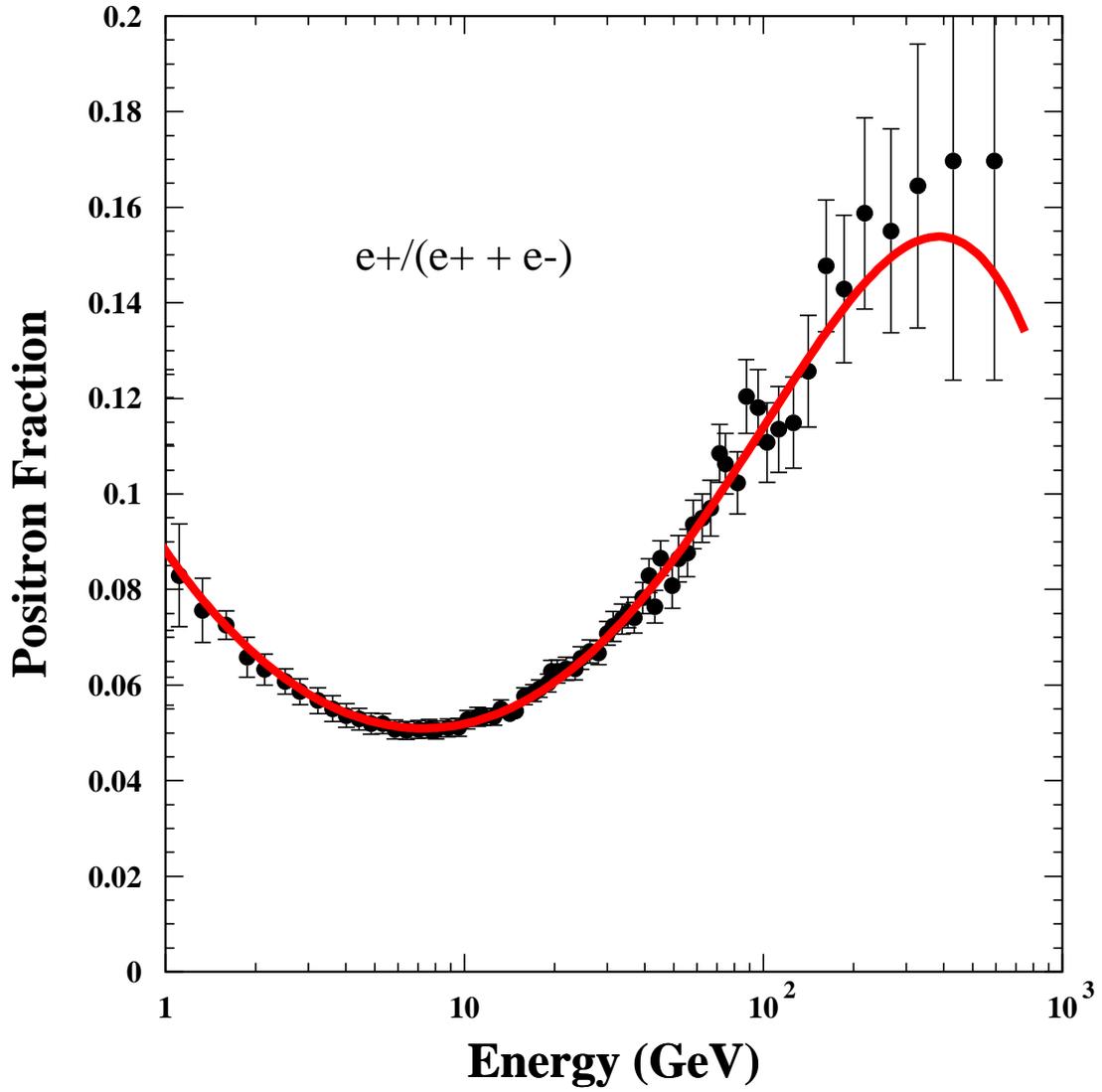,width=16.cm,height=16.cm}
\caption{The best fit 'minimal model' positron fraction 
(Accardo et al.~2014) to  the positron 
fraction measured with AMS02.}  
\label{Fig5}
\end{figure}

\begin{figure}[]
\centering
\epsfig{file=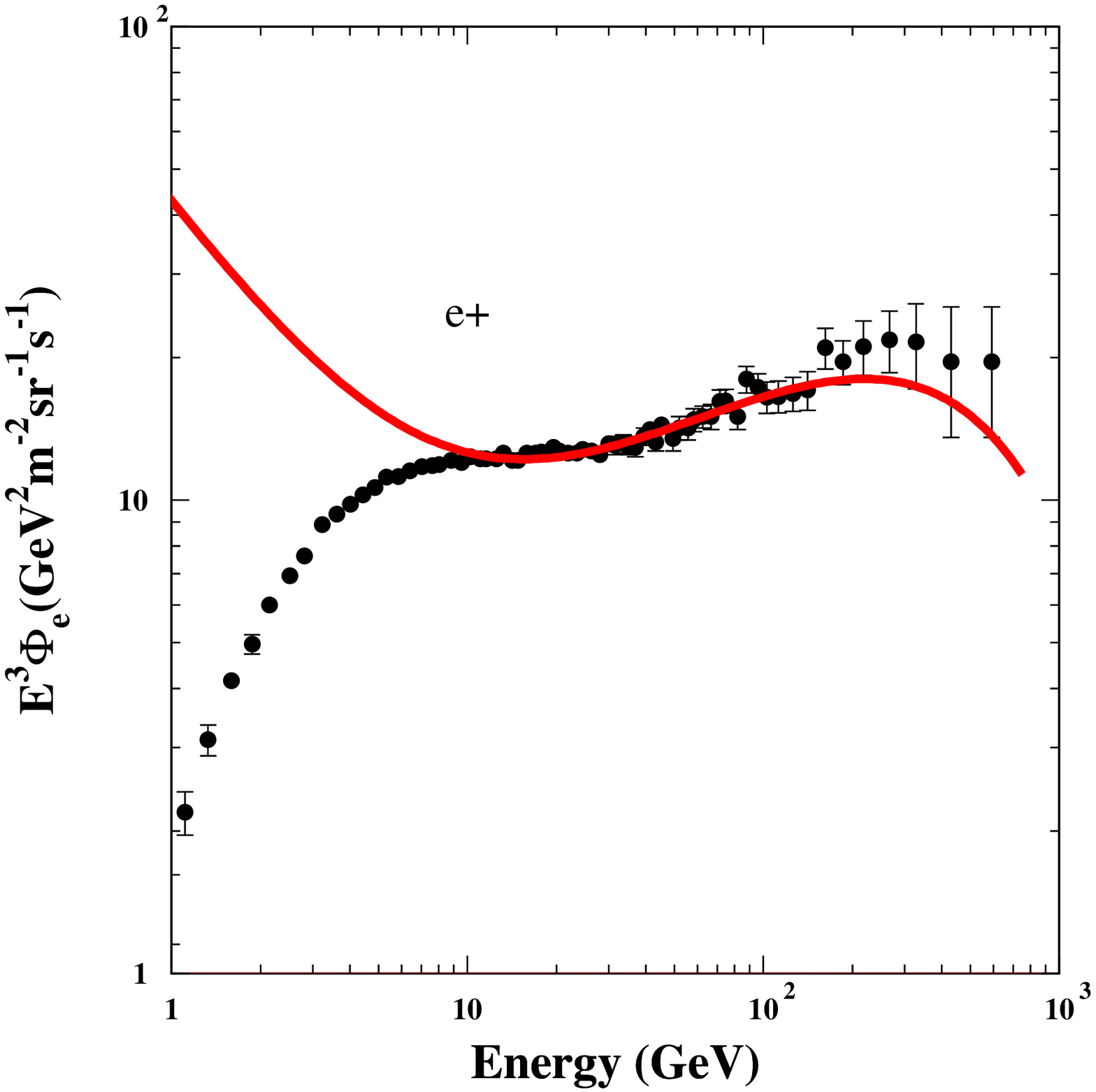,width=16.cm,height=16.cm}
\caption{Comparison between the AMS minimal model 
(Accardo et al.~2014) best fit
to the ${\rm e^+}$ CR flux, Eq.(13), and the ${\rm e^+}$ CR flux
measured with AMS02 (Aguilar et al.~ 2014).}
\label{Fig6}  
\end{figure}

\begin{figure}[]
\centering
\epsfig{file=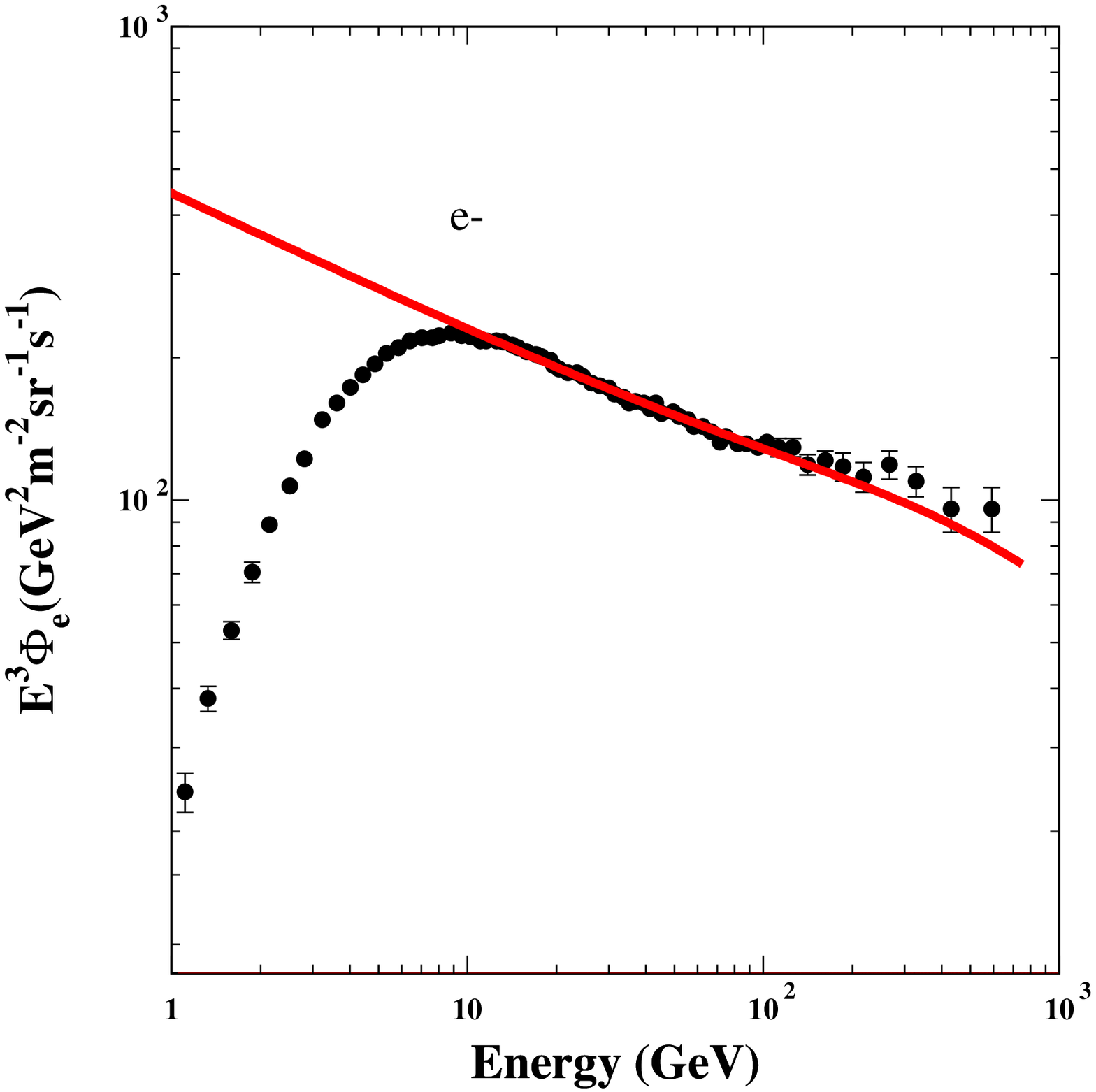,width=16.cm,height=16.cm}
\caption{Comparison between the 'AMS minimal model'     
(Accardo et al.~2014) best fit 
to the ${\rm e^-}$ CR flux, Eq.(14), and the ${\rm e^-}$ CR flux     
measured with AMS02 (Aguilar et al.~2014).} 
\label{Fig7}
\end{figure}

\begin{figure}[]
\centering
\epsfig{file=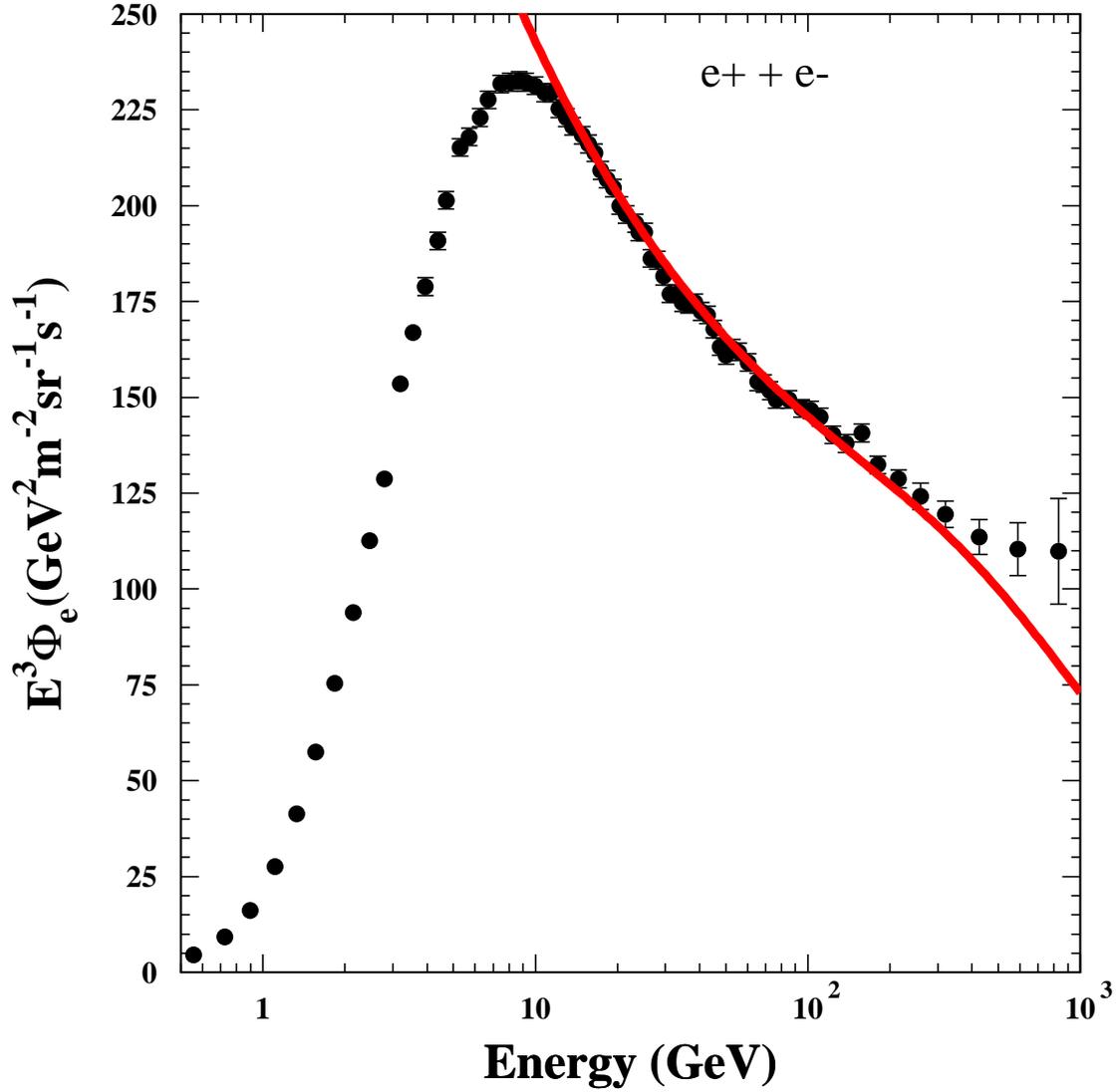,width=16.cm,height=16.cm}
\caption{Comparison between the combined 
${\rm e^\pm}$ CR flux 
measured with AMS02 near Earth  (Aguilar et al.~2014) and the flux 
obtained from the best fits of the 'AMS minimal model' (Accardo et al.~2014) 
to their measured ${\rm e^+}$ and ${\rm e^-}$ fluxes.}
\label{Fig8}
\end{figure}

\begin{figure}[]
\centering
\epsfig{file=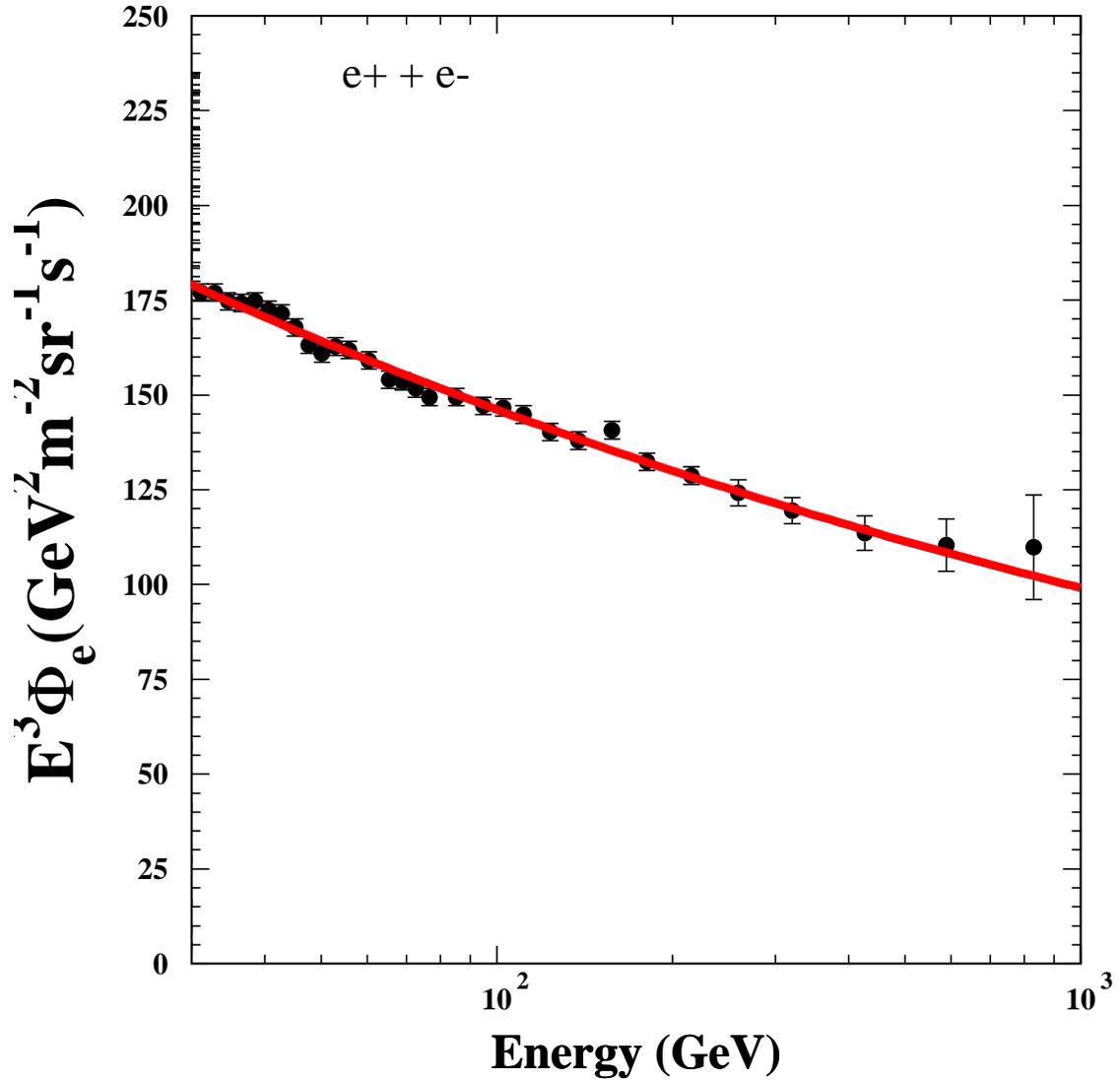,width=16.cm,height=16.cm}
\caption{A power-law ${\rm A\, E^{-3.17}}$ fit 
to the combined high energy ${\rm e^\pm}$ flux between 30-1000 GeV 
measured near Earth with AMS02 (Aguilar et al.~2014).} 
\label{Fig9}
\end{figure}

\begin{figure}[]
\centering
\epsfig{file=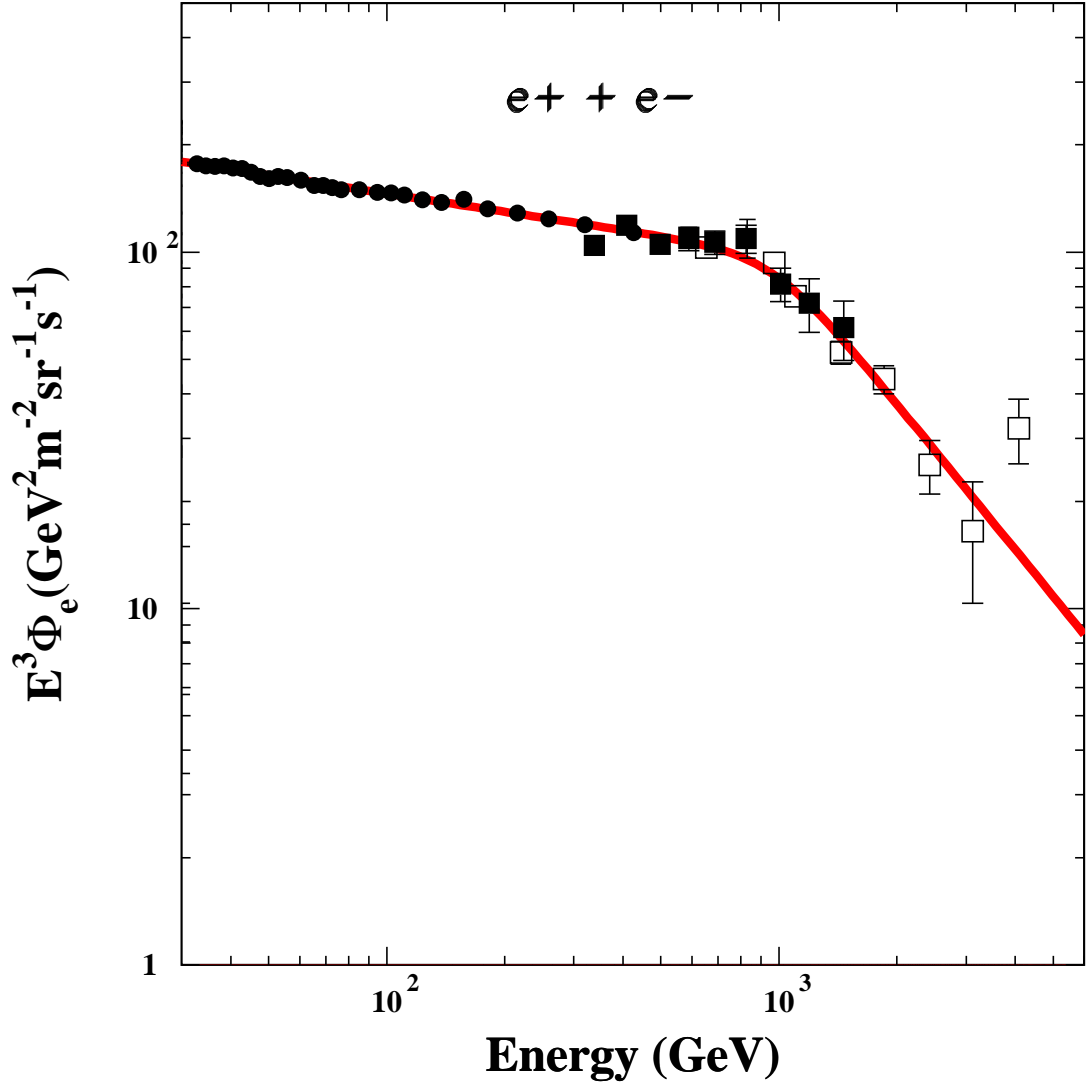,width=16.cm,height=16.cm}
\caption{A cutoff
power-law fit to the combined high energy ${\rm e^\pm}$ flux measured 
near Earth with AMS (full circles, Aguilar et al.~2014) and with H.E.S.S 
(squares, Aharonian et al.~2008,2009). The normalization of the H.E.S.S 
data was adjusted within their estimated systematic error to match
the more precise AMS02  data below TeV (Aguilar et al.~2014).}
\label{Fig10}
\end{figure}

\end{document}